\documentclass[final,1p,times]{elsarticle}

\usepackage{amssymb}
\usepackage{amsthm}
\usepackage{amsmath}
\usepackage{bm}%
\usepackage{graphics}%
\usepackage{graphicx}%
\usepackage{epsfig}%
\usepackage{subfig}
\usepackage{multirow}
\usepackage{float}
\usepackage{dcolumn}
\usepackage{upgreek}
\usepackage{setspace}
\biboptions{sort&compress}

\usepackage[nodots]{numcompress}

\journal{Elsevier}

\begin{document}

\begin{frontmatter}



\title{Strong interaction between graphene and localized hot spots in all-dielectric metasurfaces}


\author[ncu_ias]{Shuyuan Xiao\corref{cor1}}
\author[hue_phys]{Tingting Liu}
\author[hust_wnlo]{Chaobiao Zhou}
\author[hust_wnlo]{Xiaoyun Jiang}
\author[hust_wnlo]{Le Cheng}
\author[sysu_seit]{Yuebo Liu}
\author[ua_mint,ua_phys]{Zhong Li}

\address[ncu_ias]{Institute for Advanced Study, Nanchang University, Nanchang 330031, People's Republic of China}
\address[hue_phys]{Laboratory of Millimeter Wave and Terahertz Technology, School of Physics and Electronics Information, Hubei University of Education, Wuhan 430205, People's Republic of China}
\address[hust_wnlo]{Wuhan National Laboratory for Optoelectronics, Huazhong University of Science and Technology, Wuhan 430074, People's Republic of China}
\address[sysu_seit]{School of Electronics and Information Technology, Sun Yat-sen University, Guangzhou 510006, People's Republic of China}
\address[ua_mint]{Center for Materials for Information Technology, The University of Alabama, Tuscaloosa 35487, United States of America}
\address[ua_phys]{Department of Physics and Astronomy, The University of Alabama, Tuscaloosa 35487, United States of America}

\cortext[cor1]{Corresponding author. E-mail: syxiao@ncu.edu.cn (Shuyuan Xiao)}

\begin{abstract}
The active photonics based on the two-dimensional material graphene has attracted enormous interests for developing the tunable and compact optical devices with high efficiency. Here we integrate graphene into the Fano-resonant all-dielectric metasurfaces consisting of silicon split resonators, and systematically investigate the strong interaction between graphene and the highly localized hot inside feed gaps in the near infrared regime. The numerical results show that the integrated graphene can substantially reduce the Fano resonance due to the coupling effect between the intrinsic absorption of graphene with enhanced electric field in the localized hotspot. With the manipulation of the surface conductivity via varying Fermi level and the layer number of graphene, the Fano resonance strength obtains a significant modulation and is even switched off. This works provides a great degree of freedom to tailor light-matter interaction at the nanoscale and opens up the avenues for actively tunable and integrated nanophotonic device applications, such as the optical biosensing, slow light and enhanced nonlinear effects.
\end{abstract}




\end{frontmatter}


\section{Introduction}\label{sec1}
The confinement of electromagnetic radiation within nanometresized scale associated with of strong localized electromagnetic field enhancement is one of the most important and fundamental strategies in light-matter interaction engineering.\cite{gramotnev2010plasmonics} The ability of plasmonic metasurfaces to break the diffraction limit and concentrate the light into subwavelength volumes provides unprecedented opportunities to convert optical radiation into intense, localized field distributions. The typical approach to generate and control the localized field enhancement is the excitation of Fano resonances, as suggested in a variety of plasmonic metasurfaces composed of the planar arranged nanostructured metallic building blocks.\cite{luk2010fano,zhang2013multiple,moritake2014experimental,zhao2015fano,moritake2016demonstration,lim2017near} However, the high intrinsic losses and the irreversible photothermal conversion process of the traditional metal resonators have always been a challenge limiting the quality-factor ($Q$-factor) in practical optical devices. As an alternative to metal, the high-index low-loss dielectric materials, such as silicon, germanium and tellurium, have recently risen to prominence in the nanophotonics toolkit.\cite{jahani2016all,baranov2017all} The Mie resonance behavior of dielectric enables the realization of low-loss functional devices with high efficiency and exotic functionalities. Similar with the plasmonic resonance in metallic metasurfaces, the all-dielectric metasurfaces support strong Fano resonance and electromagnetic field enhancement in the optical and near-infrared spectral range.\cite{miroshnichenko2012fano,zhang2013near,zhang2017magnetically,tuz2018high,cui2018multiple} Unlike metallic metasurfaces with concentrated electromagnetic fields in the surrounding medium, the fields in dielectric metasurfaces are confined within the resonators, severely hindering the practical applications. The solutions have been suggested to introduce the feed gaps in the dielectric resonant nanostructures, which would confine larger portion of electromagnetic energy into nanoscale hot spots within the gap.\cite{novotny2011antennas,yang2014all,zhang2014strong,sun2017q,ding2018review} The feed-gapped dielectric metasurfaces serve as an ideal platform for further enhancing the interaction between light and the surrounding medium, making them promising candidates for highly sensitive optical biosensors, slow light, high-order harmonics generation, surface-enhanced enhanced Raman scattering and luminescence enhancement of quantum dots.

Recently, a new popular tendency in the research field of metasurfaces is the realization of actively controllable resonance responses, which would add more degrees of freedom in tailoring light for the development of the tunable, reconfigurable and ultracompact optical devices.\cite{zheludev2012metamaterials} On this issue, a vast amount of tuning and switching schemes have been demonstrated in plasmonic structures of metallic metasurfaces, including the mechanical reconfiguration,\cite{zhu2011switchable,ou2011reconfigurable} and the combination with phase-change materials or photosensitive materials.\cite{gu2012active,gholipour2013all,xu2016frequency,fan2017electromagnetic} Subsequently, some of these modulation strategies have been extended to dielectric metasurfaces, for instance, the combination of the all-dielectric metasurfaces with optomechanical system,\cite{zhang2013nonlinear} the temperature dependent refractive-index change of animatic liquid crystal or silicon material.\cite{zhang2014thermally,iyer2015reconfigurable,rahmani2017reversible,forouzmand2018tunable} Nevertheless, the modulation time constants in these strategies are typically on the order of 10 us to 1 ms, setting an obstacle for the real-time modulation.  Very recently, a newly emerging two-dimensional (2D) material, graphene, has attracted considerable attentions in dynamically controlling resonance response of metasurfaces, since the surface conductivity of graphene is actively tunable via shifting the Fermi level under an external bias voltage within a wide regime.\cite{ju2011graphene,he2015tunable,li2015dual,xia2018plasmonically,farmani2018broadly,li2019investigation} Moreover, the relaxation time of the excited carriers in graphene is on the order of picosecond.\cite{yee2011ultrafast,li2014ultrafast} These exceptional properties makes graphene an excellent competitor than other active materials to be combined with metasurfaces for ultrafast control of optical resonances.\cite{li2016monolayer,xiao2017strong,chen2017study,xiao2018active,kim2018electrically,li2019terahertz} On this respect, researchers have been inspired to incorporate graphene with all-dielectric metasurfaces for active tunability. For instance, the combination of dielectric metasurfaces with graphene layer is proposed to achieve the tunable transmission of the trapped modes for optical modulator.\cite{argyropoulos2015enhanced,chen2017flatland} However, in these pioneering works, the field enhancements are confined inside the dielectric resonators, which limits the interaction with graphene and degrades the efficiency of the functional devices to an extent.

In this work, we provide a systematic investigation on the interaction between graphene with the localized hot spots based on feed-gapped all-dielectric metasurfaces in near infrared regime. The unique feature of the metasurfaces consisting of a pair of asymmetric split silicon bars is that the dielectric resonators support extremely high $Q$-factor Fano resonance and generate strong localized hot spots within the gaps. With graphene covered over the metasurfaces, the Fano resonance strength exhibits a remarkable modulation due to the strong interaction of the enhanced electric field with graphene. The interaction effect is detailedly investigated by changing the gap width, manipulating the surface conductivity via varying Fermi level and the layer number of graphene. To the best of our knowledge, this is the first investigation of the interaction between graphene and localized hot spots of Fano resonance in all-dielectric metasurfaces, offering great prospects in designing actively tunable and integrated nanophotonic devices with intriguing functionalities.

\section{The geometric structure and numerical model}\label{sec2}
The schematic of the designed hybrid graphene-dielectric metasurfaces is illustrated in Figure~1(a). The unit cell of the all-dielectric metasurfaces is shown in Figure~1(b), following the classic design of feed-gapped structure to generate the hot spots, which consists of two silicon split nanobar resonators periodically patterned on a silica substrate.\cite{zhang2014strong,sun2017q} The periods along $x$ and $y$ directions are the same as 900 nm. The two asymmetric parallel silicon nanobars have the slight different lengths as 700 nm and 750 nm, and share the same width of 200 nm as well as the same thickness of 160 nm. The feed gaps are introduced here and each of the two silicon nanobars has a 25 nm wide groove in the center for the enhancement of the interaction between the fields and the surrounding medium. The graphene layer is then covered over the dielectric silicon nanobars in a continuous pattern. The dielectric particles with different geometries have been widely employed as building blocks of the metasurface due to their low intrinsic loss and especially their easy fabrication using top-down method. The graphene layer grown from the chemical vapor deposition can be transferred to the dielectric metasurfaces using standard transfer techniques.\cite{suk2011transfer}
\begin{figure}[htbp]
\centering
\includegraphics[scale=0.60]{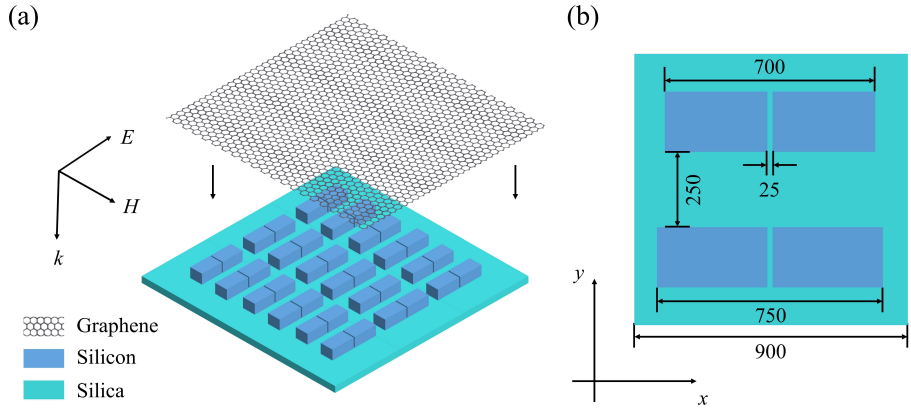}
\caption{(a) Schematic illustration of the proposed hybrid graphene-dielectric metasurfaces. (b) Top view and geometric parameters of a unit cell in the all-dielectric metasurfaces.\label{fig1}}
\end{figure}

Numerical simulations of the hybrid metasurfaces are carried out using the finite-difference time-domain (FDTD) approach based on the commercial software package \textsl{Lumerical Solutions}. In the simulations, the periodical boundary conditions are used in the $x$ and $y$ directions, and the perfectly matched layer is employed in the $z$ direction. The plane waves are incident along $-z$ direction and provide an electric field polarized along the $x$ direction. The silicon and the silica are assumed to be lossless with the constant refractive index of $n_{\text{Si}}=3.5$ and $n_{\text{SiO2}}=1.4$.\cite{palik1985handbook} The graphene over the all-dielectric metasurfaces can be modeled as a 2D sheet and the surface conductivity is governed by the random phase approximation (RPA) in the local limit. The conductivity of graphene is related to the Fermi level $E_{\text{F}}$ and includes the interband and intraband contributions as follows,\cite{zhang2015towards,xiao2016tunable}
\begin{equation}
    \begin{split}
      \sigma_{\text{g}} &=\sigma_{\text{intra}}+\sigma_{\text{inter}}=\frac{2e^{2}k_{\text{B}}T}{\pi\hbar^{2}}\frac{i}{\omega+i\tau^{-1}}\ln[2\cosh(\frac{E_{\text{F}}}{2k_{\text{B}}T})]\\
                        &+\frac{e^2}{4\hbar}[\frac{1}{2}+\frac{1}{\pi}\arctan(\frac{\hbar\omega-2E_{\text{F}}}{2k_{\text{B}}T}) \\
                        &-\frac{i}{2\pi}\ln\frac{(\hbar\omega+2E_{\text{F}})^{2}}{(\hbar\omega-2E_{\text{F}})^{2}+4(k_{\text{B}}T)^{2}}],\label{eq1}
    \end{split}
\end{equation}
where $e$ is the electron charge, $k_{\text{B}}$ is the Boltzmann constant, $T$ is the environment temperature with constant value of 300 K here, $\hbar$ is the reduced Planck's constant, $\omega$ is the incident light frequency, and $\tau$ is the carrier relaxation time. $\tau$ is calculated from $\tau=(\mu E_{\text{F}})/(e v_{\text{F}}^{2})$ and depends on the carrier mobility $\mu$, the Fermi level $E_{\text{F}}$ and the Fermi velocity $v_{\text{F}}$. In accordance with the experimental results in \cite{kim2012electrical}, we set $\mu=10000$ cm$^{2}$/V$\cdot$s and $v_{\text{F}}=1\times 10^{6}$ m/s in calculating the conductivity of the doped graphene. Therefore, the optical conductivity of the graphene can obtain a continuous manipulation ($0\sim e^{2}/4\hbar$) by changing the Fermi level, as depicted below. The Figure~2(a) and (b) plot respective variation of the real part and imaginary part of the graphene conductivity as incident light frequency and Fermi level increases. Because the interband contribution to graphene conductivity will be blocked once the Fermi level exceeds the Dirac point by half of the photon energy ($E_{\text{F}}>\hbar\omega/2$), the real part of the conductivity will rapidly reduce to around zero when the Fermi level is larger than the critical value. Hence, a clear boundary line can be observed in Figure~2(a). On the contrary, the imaginary part of graphene conductivity in Figure~2(b) shows a continuous increase due to the significant contributions of the intraband transition as Fermi level increases across the critical value. The unique property of the graphene conductivity enables the flexible tunability of the localized hot spots in the near infrared regime.
\begin{figure}[htbp]
\centering
\includegraphics[scale=0.60]{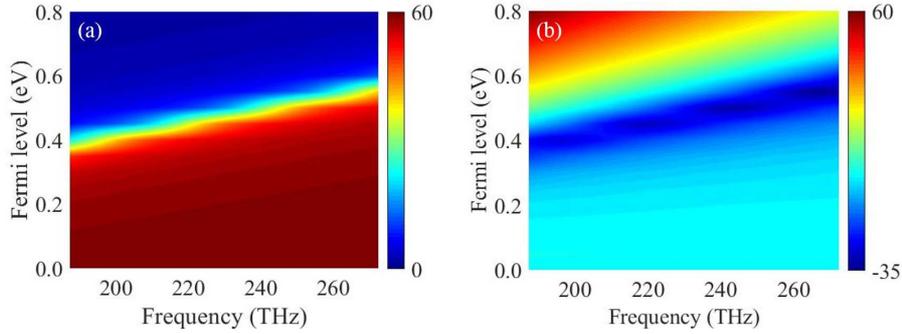}
\caption{The variations of (a) real and (b) imaginary part of the graphene conductivity as incident light frequency and Fermi level increases.\label{fig2}}
\end{figure}

\section{Simulation results and discussions}\label{sec3}
To begin with, the transmission spectra of the all-dielectric metasurfaces composed of silicon split nanobars are presented in Figure~3(a). Without the presence of the graphene layer, one can clearly observe a sharp asymmetric Fano dip at 213.36 THz in the regime of interest. Due to the reduced intrinsic losses of silicon material, the Fano resonance of the structure exhibits a high $Q$-factor. The $Q$-factor in a typical Fano resonance can be obtained through fitting the Fano line shape by the following formula,\cite{wu2014spectrally}
\begin{equation}
    T=|a_{1}+ia_{2}+\frac{b}{\omega-\omega_{0}+i\gamma}|^{2},\label{eq2}
\end{equation}
where $a_{1}$, $a_{2}$ and $b$ are in general real constants for fitting, $\omega_{0}$ is the central resonance frequency and $\gamma$ is the damping rate of the resonance. The $Q$-factor is then estimated from $Q=\omega_{0}/(2\gamma)$ as 1190.79.

The distribution of the corresponding electric field $E/E_{\text{0}}$ at the dip of the transmission spectrum is presented in Figure~3(b). The Fano resonance results from the anti-phased oscillation of displacement currents of the two split nanobar resonators with accumulated polarization charges. Especially, a large amount of opposite charges has been confined at the two sides of the gaps, leading to a large portion of electromagnetic energy density localized at the gap and finally a significant field enhancement within the gaps. The introduction of the feed gaps at the center of the silicon nanobars enables the further enhancement of the localized hotspot inside the gap. Consequently, the high $Q$-factor and the strong localized hotspot in the designed all-dielectric metasurface allows to enhance the interactions between light and the surrounding medium.
\begin{figure}[htbp]
\centering
\includegraphics[scale=0.60]{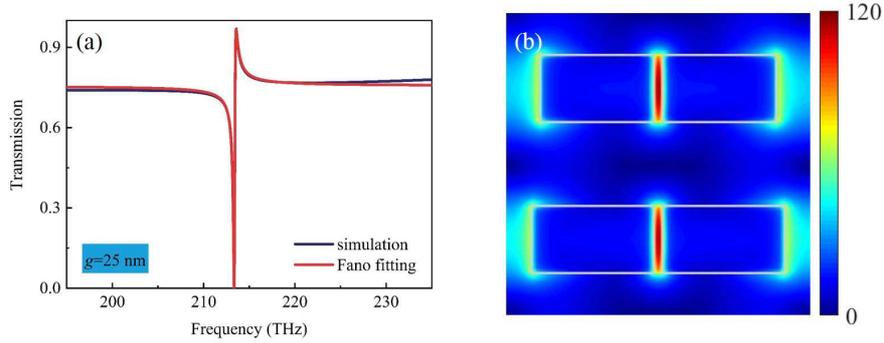}
\caption{(a) Simulated and fitted transmission spectrum of the all-dielectric metasurface without the graphene layer. (b) Distribution of the corresponding electric field $E/E_{\text{0}}$ at the resonance frequency of 213.36 THz.\label{fig3}}
\end{figure}

Then the Fano resonance characteristics of the structure are investigated for the two cases with and without the graphene layer on different feed-gapped nanobar arrays, as shown in Figure~4(a)-(d). Without the presence of the graphene layer, the asymmetric Fano shapes can be observed in transmission spectra for different gapped nanobar arrays. The variations of the feed gap widths in silicon nanobars of the metasurface provide the freedom to engineer the asymmetry degrees of the unit cell in the $y$ direction, giving rise to the modulation of the high $Q$-factor resonances with significant localized hot spots inside the gaps. As the gap width of the silicon nanobars increase from 25 to 200 nm, the resonance frequency shows a blue shift since the increase of feed gap width leads to the reduction of the effective length of the silicon nanobars. The $Q$-factor of the resonances gradually decreases from 1190.79, 1129.78, 1008.17 to 957.99 as the gap width increases from 25, 50, 100 to 200 nm, arising from the increasing radiative losses. In addition, the electric field distributions of the different gapped dielectric metasurfaces are plotted in Figure 5(a)-(d). As the gap of the silicon nanobars becomes wider, the electric fields concentrated at ends of the resonators become gradually weaker and the accumulation of the opposite charges at the two sides of the gaps shows a declining tendency. The decreasing proportion of the electromagnetic energy results in the smaller localized hotspot amplitude.
\begin{figure}[htbp]
\centering
\includegraphics[scale=0.60]{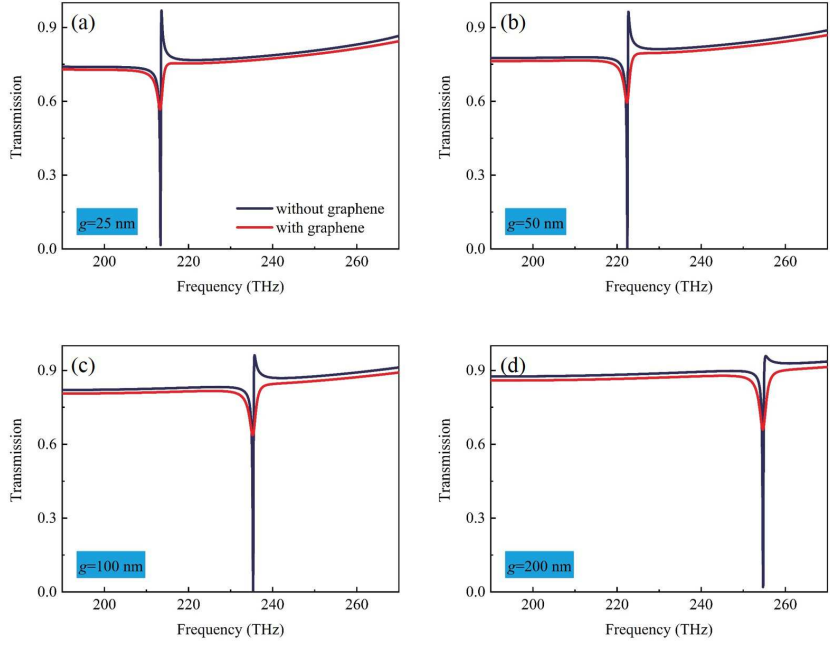}
\caption{The simulated transmission spectra of the dielectric metasurfaces with and without the graphene layer in silicon nanobar arrays with different gap widths.\label{fig4}}
\end{figure}

With the presence of graphene covered over the dielectric metasurfaces, though the trend of the blue shift in the the resonance frequency is exactly retained, the resonance strength of Fano resonance undergoes a remarkable modulation, which indicates the strong interaction between graphene and the localized hot spots. As shown in Figure~4(a)-(d), the transmission amplitude of the resonance rapidly decreases when the undoped graphene layer with a fixed Fermi level $E_{\text{F}}=0$ eV (the optical conductivity of $e^{2}/4\hbar$) is placed on the silicon nanobar arrays. To character the amplitude modulation in a quantitative way, we introduce the absolute value of modulation depth as $\Delta T=|T_{\text{g}}-T_{\text{0}}|\times100\%$, where $T_{\text{g}}$ and $T_{\text{0}}$ are the transmission coefficients at the resonance dip for the dielectric metasurfaces with and without graphene, respectively. The modulation depth can be calculated as $\Delta T=55.08\%$, $58.96\%$, $63.32\%$, $64.03\%$ for the gap width of 25, 50, 100, 200 nm, respectively.
\begin{figure}[htbp]
\centering
\includegraphics[scale=0.70]{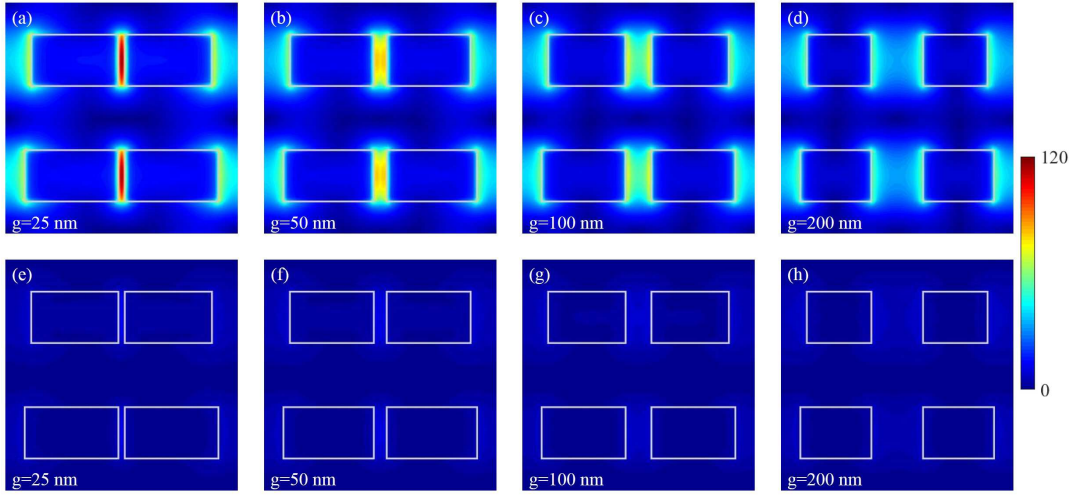}
\caption{The distributions of the electric field $E/E_{\text{0}}$ at the resonance frequency for silicon nanobar arrays with different gap widths (e)-(h) without the graphene layer and (i)-(l) with the presence of the graphene layer.\label{fig5}}
\end{figure}

To understand the interaction mechanism, the corresponding electric field distributions at the Fano resonances with the graphene layer are compared in Figure~5(e)-(h) to reveal the coupling effect between graphene layer and the localized hot spots. As mentioned above, the high $Q$-factor and the strong localized hot spots provide a platform for the enhanced interaction between light and the surrounding medium. As the only lossy material in the hybrid structure, the monolayer graphene strongly couples to the incident light with its intrinsic absorption once it is integrated into the all-dielectric metasurfaces. According to Eq. (1), the interband contribution is dominant in the conductivity at near infrared frequencies, and the high conductivity of graphene leads to a remarkable light absorption. As a result, the electric field within the feed gaps dramatically decreases, leading to the shrunk outline of the localized hot spots and the reduction in Fano resonance strength, which corresponds to the significant change in the transmission coefficients. By contrast, we also present the modulation of the undoped graphene on the strength of Fano resonance in the dielectric metasurfaces consisting of the continuous nanobars without feed gaps (not shown). The modulation depth of only $48\%$ is much lower than that of the feed-gapped structure due to the severely limited interaction between graphene and the electric field inside the continuous dielectric nanobars.

In Figure~6, the dependences of the maximum field enhancement in the feed gaps and the modulation depth upon the feed gap width are summarized for the designed metasurfaces. The maximum field enhancement factor gradually decreases from 120 to 55 with increasing gap width from 25 to 200 nm, because the coupling effect between the two parts of the split bar becomes weaker and the accumulation of the opposite charges at the two sides of the gap gradually reduces with larger gap width. In contrast to the decreasing electric field enhancement, the modulation depth $\Delta T$ of the transmission strength increases from $55\%$ to $64\%$ in the hybrid graphene-dielectric metasurfaces, implying even stronger interaction between graphene and the localized hot spots. Such an opposite tendency can be attributed to the fact that the volume for the interaction becomes larger with larger gap width and a larger proportion of electromagnetic field in the gap would interact with the graphene. Hence the volume plays a dominant role rather than the field enhancement in the evaluation of the interaction strength. In addition, the modulation depth $\Delta T$ shows the saturation tendency as gap width increases to more than 150 nm, since the proportion of the electromagnetic field in the gap would not continue to increase resulting from the counteraction between the weaker field enhancement and the larger gap width. Hence, although the field enhancement of the dielectric structure can be further increased with a shorter gap, which exactly satisfies the requirements of some applications including nonlinear optics and enhanced Raman scattering, in some other specific applications such as optical modulation (in the present situation) as well as biosensing, the volume for the interaction between electromagnetic fields and the surrounding medium is more significant. In this case, the optimum width of the feed gap needs to be considered under the modulation depth limit for the reconfigurable and compact devices.
\begin{figure}[htbp]
\centering
\includegraphics[scale=0.40]{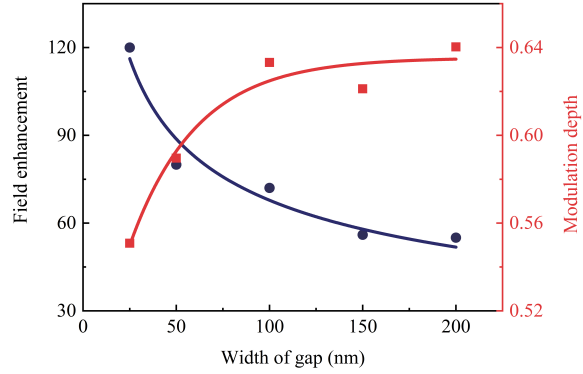}
\caption{The dependence of the maximum field enhancement $E/E_{\text{0}}$ and the modulation depth $\Delta T$ of the metasurface on the feed gap width. The dots are the calculated values and the lines are the fitted curves.\label{fig6}}
\end{figure}

Next we investigate the modulation potential of monolayer graphene arising from the continuously tunable conductivity via shifting the Fermi level. In the hybrid metasurfaces here, the feed gap of the silicon nanobars is fixed as 25 nm in width, and the Fermi level of graphene is varied as $E_{\text{F}}=0$ eV, 0.5 eV and 0.6 eV. The transmission spectra of the metasurfaces under different Fermi levels of graphene are calculated in Figure~7. Compared with the pronounced dip in the transmission spectrum of all-dielectric metasurfaces without graphene, the amplitude modulation of the asymmetric Fano transmission spectra can be observed with the Fermi level $E_{\text{F}}=0.6$ eV, and much further stronger with $E_{\text{F}}=0.5$ eV. Finally, the sharp resonance spectrum becomes broad and flat for the case of the undoped graphene with $E_{\text{F}}=0$ eV. The absolute value of the modulation depth $\Delta T$ is employed to evaluate the modulation performance. With the results of $\Delta T=55.08\%$, $14.79\%$, $0.17\%$ for $E_{\text{F}}=0$ eV, 0.5 eV and 0.6 eV, the modulation depth gradually decreases when the Fermi level of graphene increases. By shifting the Fermi level of the graphene layer in the structure, the transmission amplitude of Fano resonance is dynamically controllable with a nearly fixed resonance frequency.
\begin{figure}[htbp]
\centering
\includegraphics[scale=0.40]{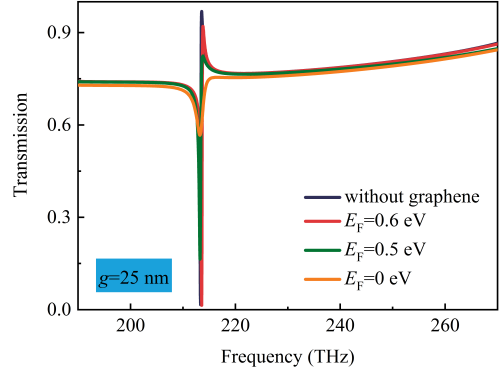}
\caption{The dependence of the transmission spectra of the hybrid metasurfaces on the Fermi level of graphene.\label{fig7}}
\end{figure}

The efficient modulation in the transmission strength of the metasurface originates from the strong interaction between the graphene layer and the localized hot spots in the gaps, to be exactly, arises from the absorption effect of graphene. In the near infrared regime of interest, the real part of the graphene conductivity decreases with the increase of the Fermi level from 0 eV to 0.6 eV, as shown in Figure~2 (a). Especially, when Fermi level increases to the critical point, calculated as 0.44 eV at the resonance frequency of 213.36 THz, the contribution of interband transition to conductivity rapidly decreases due to the optical Pauli blocking. Accordingly, the real part of conductivity shows a remarkable reduction with increasing the Fermi level, leading to the reduced light absorption and the weak tunability of the resonance strength. Thus, the localized hot spots in the structure can be actively switchable by manipulating the interaction between the graphene absorption and enhanced electric field.

Considering the inevitable disorders during the synthesis or transfer processes of the graphene layer in practical fabrication, the influence of the multilayer graphene on the Fano resonance is also discussed here. According to the investigations in \cite{hass2008multilayer}, the randomly stacked graphene layers behaves as the isolated monolayer graphene due to the electrical decoupling, and the conductivity is proportional to the layer number. Combined with the influence of conductivity of monolayer graphene on resonance strength of the metasurface, the structure covered with multilayer graphene will further decrease the amplitudes of the transmission spectra due to the multiple conductivity. Figure~8 provides the simulated transmission spectra of the hybrid metasurface covered by different graphene layers. In this configurations, the Fermi level is set as $E_{\text{F}}=0$ eV and the feed gap in the silicon nanobar is fixed to 25 nm in width. It is observed that the transmission amplitude of the resonance shows a rapid decline as the layer number of graphene increases at a nearly fixed resonance frequency, and the resonance is even switched off as the layer number reaches 3. The modulation depth can be calculated as $\Delta T=55.08\%$, $61.82\%$, $63.66\%$, for the layer number of 1, 2 and 3, respectively. The simulation results not only demonstrates the dependence of the resonance on the layer number of graphene in the hybrid metasurfaces, but also further validates the proposed physical mechanism of modulation from the interaction between the intrinsic absorption of graphene and localized hot spots. In addition, the high sensitivity of the Fano resonance to the multilayer graphene also shows a great advantage in ultrasensitive biosensing applications. When the detected molecules is attached at the gap where the electric field is remarkably enhanced, it is expected that the absorption of the molecules leads to the modification of the polarizability of the graphene layer and thus changes the shape or the position of Fano resonance.
\begin{figure}[htbp]
\centering
\includegraphics[scale=0.40]{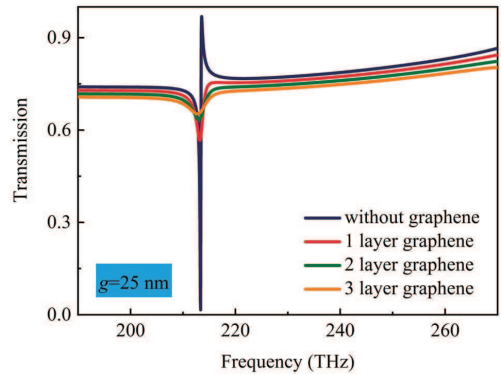}
\caption{The dependence of the transmission spectra of the hybrid metasurfaces on the layer number of graphene.\label{fig8}}
\end{figure}

\section{Conclusions}\label{sec4}
In conclusions, we symmetrically investigate the interaction between graphene with Fano resonant all-dielectric metasurfaces. Using the silicon split resonators as the building blocks of the metasurfaces, the Fano resonance with a remarkably high $Q$-factor and highly localized hot spots inside the gaps are observed. With graphene covered on the dielectric structure, the Fano resonance strength is manipulated by changing the gap width of the resonators, shifting the Fermi level of graphene and varying the layer number of graphene. The dependence of the field enhancement and the modulation depth upon the gap width of the resonators reveals out the dominant role of the volume in the interaction between graphene and dielectric structure. The modulation in the transmission amplitude of Fano resonance by varying the Fermi level and the layer number of graphene further demonstrates the strong interaction between the intrinsic absorption of graphene with the electric field in the localized hot spots. This work offering great prospects in designing active and compact metamaterial devices such as light modulators, switches and biosensors, and the compatibility of graphene with dielectric materials would endow it with remarkable applicability for integration technologies in the near future.

\section*{Acknowledgments}
This work is supported by the National Natural Science Foundation of China (Grant No. 61775064, 11847100 and 11847132), the Fundamental Research Funds for the Central Universities (HUST: 2016YXMS024) and the Natural Science Foundation of Hubei Province (Grant No. 2015CFB398 and 2015CFB502).


\begin{thebibliography}{54}
	\providecommand{\natexlab}[1]{#1}
	\providecommand{\url}[1]{\texttt{#1}}
	\providecommand{\href}[2]{#2}
	\providecommand{\path}[1]{#1}
	\providecommand{\eprint}[1]{\href{http://arxiv.org/abs/#1}{\path{#1}}}
	\providecommand{\DOIprefix}{doi:}
	\providecommand{\ArXivprefix}{arXiv:}
	\providecommand{\URLprefix}{URL: }
	\providecommand{\Pubmedprefix}{pmid:}
	\providecommand{\doi}[1]{\href{http://dx.doi.org/#1}{\path{#1}}}
	\providecommand{\Pubmed}[1]{\href{pmid:#1}{\path{#1}}}
	\providecommand{\BIBand}{and}
	\providecommand{\bibinfo}[2]{#2}
	\ifx\xfnm\undefined \def\xfnm[#1]{\unskip,\space#1}\fi
	\bibitem[{Gramotnev and Bozhevolnyi(2010)}]{gramotnev2010plasmonics}
	\bibinfo{author}{Gramotnev\xfnm[ D.K.]}, \bibinfo{author}{Bozhevolnyi\xfnm[
		S.I.]}.
	\newblock \bibinfo{title}{Plasmonics beyond the diffraction limit}.
	\newblock \bibinfo{journal}{Nat Photonics}
	\bibinfo{year}{2010};\bibinfo{volume}{4}(\bibinfo{number}{2}):\bibinfo{pages}{83}.
	\bibitem[{Luk'yanchuk et~al.(2010)Luk'yanchuk, Zheludev, Maier, Halas,
		Nordlander, Giessen et~al.}]{luk2010fano}
	\bibinfo{author}{Luk'yanchuk\xfnm[ B.]}, \bibinfo{author}{Zheludev\xfnm[
		N.I.]}, \bibinfo{author}{Maier\xfnm[ S.A.]}, \bibinfo{author}{Halas\xfnm[
		N.J.]}, \bibinfo{author}{Nordlander\xfnm[ P.]},
	\bibinfo{author}{Giessen\xfnm[ H.]}, et~al.
	\newblock \bibinfo{title}{The fano resonance in plasmonic nanostructures and
		metamaterials}.
	\newblock \bibinfo{journal}{Nat Mater}
	\bibinfo{year}{2010};\bibinfo{volume}{9}(\bibinfo{number}{9}):\bibinfo{pages}{707}.
	\bibitem[{Zhang et~al.(2013{\natexlab{a}})Zhang, Wen, Li, Ruan, Wang and
		Xiong}]{zhang2013multiple}
	\bibinfo{author}{Zhang\xfnm[ Q.]}, \bibinfo{author}{Wen\xfnm[ X.]},
	\bibinfo{author}{Li\xfnm[ G.]}, \bibinfo{author}{Ruan\xfnm[ Q.]},
	\bibinfo{author}{Wang\xfnm[ J.]}, \bibinfo{author}{Xiong\xfnm[ Q.]}.
	\newblock \bibinfo{title}{Multiple magnetic mode-based fano resonance in
		split-ring resonator/disk nanocavities}.
	\newblock \bibinfo{journal}{ACS Nano}
	\bibinfo{year}{2013}{\natexlab{a}};\bibinfo{volume}{7}(\bibinfo{number}{12}):\bibinfo{pages}{11071--11078}.
	\bibitem[{Moritake et~al.(2014)Moritake, Kanamori and
		Hane}]{moritake2014experimental}
	\bibinfo{author}{Moritake\xfnm[ Y.]}, \bibinfo{author}{Kanamori\xfnm[ Y.]},
	\bibinfo{author}{Hane\xfnm[ K.]}.
	\newblock \bibinfo{title}{Experimental demonstration of sharp fano resonance in
		optical metamaterials composed of asymmetric double bars}.
	\newblock \bibinfo{journal}{Opt Lett}
	\bibinfo{year}{2014};\bibinfo{volume}{39}(\bibinfo{number}{13}):\bibinfo{pages}{4057--4060}.
	\bibitem[{Zhao et~al.(2015)Zhao, Zhang, Zhu, Yuan and Qin}]{zhao2015fano}
	\bibinfo{author}{Zhao\xfnm[ J.]}, \bibinfo{author}{Zhang\xfnm[ J.]},
	\bibinfo{author}{Zhu\xfnm[ Z.]}, \bibinfo{author}{Yuan\xfnm[ X.]},
	\bibinfo{author}{Qin\xfnm[ S.]}.
	\newblock \bibinfo{title}{Fano resonances and strong field enhancements in
		arrays of asymmetric plasmonic gap-antennas}.
	\newblock \bibinfo{journal}{J Opt}
	\bibinfo{year}{2015};\bibinfo{volume}{17}(\bibinfo{number}{8}):\bibinfo{pages}{085002}.
	\bibitem[{Moritake et~al.(2016)Moritake, Kanamori and
		Hane}]{moritake2016demonstration}
	\bibinfo{author}{Moritake\xfnm[ Y.]}, \bibinfo{author}{Kanamori\xfnm[ Y.]},
	\bibinfo{author}{Hane\xfnm[ K.]}.
	\newblock \bibinfo{title}{Demonstration of sharp multiple fano resonances in
		optical metamaterials}.
	\newblock \bibinfo{journal}{Opt Express}
	\bibinfo{year}{2016};\bibinfo{volume}{24}(\bibinfo{number}{9}):\bibinfo{pages}{9332--9339}.
	\bibitem[{Lim et~al.(2017)Lim, Han, Gupta, MacDonald and Singh}]{lim2017near}
	\bibinfo{author}{Lim\xfnm[ W.X.]}, \bibinfo{author}{Han\xfnm[ S.]},
	\bibinfo{author}{Gupta\xfnm[ M.]}, \bibinfo{author}{MacDonald\xfnm[ K.F.]},
	\bibinfo{author}{Singh\xfnm[ R.]}.
	\newblock \bibinfo{title}{Near-infrared linewidth narrowing in plasmonic
		fano-resonant metamaterials via tuning of multipole contributions}.
	\newblock \bibinfo{journal}{Appl Phys Lett}
	\bibinfo{year}{2017};\bibinfo{volume}{111}(\bibinfo{number}{6}):\bibinfo{pages}{061104}.
	\bibitem[{Jahani and Jacob(2016)}]{jahani2016all}
	\bibinfo{author}{Jahani\xfnm[ S.]}, \bibinfo{author}{Jacob\xfnm[ Z.]}.
	\newblock \bibinfo{title}{All-dielectric metamaterials}.
	\newblock \bibinfo{journal}{Nat Nanotechnol}
	\bibinfo{year}{2016};\bibinfo{volume}{11}(\bibinfo{number}{1}):\bibinfo{pages}{23}.
	\bibitem[{Baranov et~al.(2017)Baranov, Zuev, Lepeshov, Kotov, Krasnok,
		Evlyukhin et~al.}]{baranov2017all}
	\bibinfo{author}{Baranov\xfnm[ D.G.]}, \bibinfo{author}{Zuev\xfnm[ D.A.]},
	\bibinfo{author}{Lepeshov\xfnm[ S.I.]}, \bibinfo{author}{Kotov\xfnm[ O.V.]},
	\bibinfo{author}{Krasnok\xfnm[ A.E.]}, \bibinfo{author}{Evlyukhin\xfnm[
		A.B.]}, et~al.
	\newblock \bibinfo{title}{All-dielectric nanophotonics: the quest for better
		materials and fabrication techniques}.
	\newblock \bibinfo{journal}{Optica}
	\bibinfo{year}{2017};\bibinfo{volume}{4}(\bibinfo{number}{7}):\bibinfo{pages}{814--825}.
	\bibitem[{Miroshnichenko and Kivshar(2012)}]{miroshnichenko2012fano}
	\bibinfo{author}{Miroshnichenko\xfnm[ A.E.]}, \bibinfo{author}{Kivshar\xfnm[
		Y.S.]}.
	\newblock \bibinfo{title}{Fano resonances in all-dielectric oligomers}.
	\newblock \bibinfo{journal}{Nano Lett}
	\bibinfo{year}{2012};\bibinfo{volume}{12}(\bibinfo{number}{12}):\bibinfo{pages}{6459--6463}.
	\bibitem[{Zhang et~al.(2013{\natexlab{b}})Zhang, MacDonald and
		Zheludev}]{zhang2013near}
	\bibinfo{author}{Zhang\xfnm[ J.]}, \bibinfo{author}{MacDonald\xfnm[ K.F.]},
	\bibinfo{author}{Zheludev\xfnm[ N.I.]}.
	\newblock \bibinfo{title}{Near-infrared trapped mode magnetic resonance in an
		all-dielectric metamaterial}.
	\newblock \bibinfo{journal}{Opt Express}
	\bibinfo{year}{2013}{\natexlab{b}};\bibinfo{volume}{21}(\bibinfo{number}{22}):\bibinfo{pages}{26721--26728}.
	\bibitem[{Zhang et~al.(2017)Zhang, Li, He, Chen, Fan, Zhao
		et~al.}]{zhang2017magnetically}
	\bibinfo{author}{Zhang\xfnm[ F.]}, \bibinfo{author}{Li\xfnm[ C.]},
	\bibinfo{author}{He\xfnm[ X.]}, \bibinfo{author}{Chen\xfnm[ L.]},
	\bibinfo{author}{Fan\xfnm[ Y.]}, \bibinfo{author}{Zhao\xfnm[ Q.]}, et~al.
	\newblock \bibinfo{title}{Magnetically coupled fano resonance of dielectric
		pentamer oligomer}.
	\newblock \bibinfo{journal}{J Phys D: Appl Phys}
	\bibinfo{year}{2017};\bibinfo{volume}{50}(\bibinfo{number}{27}):\bibinfo{pages}{275002}.
	\bibitem[{Tuz et~al.(2018)Tuz, Khardikov, Kupriianov, Domina, Xu, Wang
		et~al.}]{tuz2018high}
	\bibinfo{author}{Tuz\xfnm[ V.R.]}, \bibinfo{author}{Khardikov\xfnm[ V.V.]},
	\bibinfo{author}{Kupriianov\xfnm[ A.S.]}, \bibinfo{author}{Domina\xfnm[
		K.L.]}, \bibinfo{author}{Xu\xfnm[ S.]}, \bibinfo{author}{Wang\xfnm[ H.]},
	et~al.
	\newblock \bibinfo{title}{High-quality trapped modes in all-dielectric
		metamaterials}.
	\newblock \bibinfo{journal}{Opt Express}
	\bibinfo{year}{2018};\bibinfo{volume}{26}(\bibinfo{number}{3}):\bibinfo{pages}{2905--2916}.
	\bibitem[{Cui et~al.(2018)Cui, Zhou, Yuan, Qiu, Zhu, Wang
		et~al.}]{cui2018multiple}
	\bibinfo{author}{Cui\xfnm[ C.]}, \bibinfo{author}{Zhou\xfnm[ C.]},
	\bibinfo{author}{Yuan\xfnm[ S.]}, \bibinfo{author}{Qiu\xfnm[ X.]},
	\bibinfo{author}{Zhu\xfnm[ L.]}, \bibinfo{author}{Wang\xfnm[ Y.]}, et~al.
	\newblock \bibinfo{title}{Multiple fano resonances in symmetry-breaking silicon
		metasurface for manipulating light emission}.
	\newblock \bibinfo{journal}{ACS Photonics}
	\bibinfo{year}{2018};\bibinfo{volume}{5}(\bibinfo{number}{10}):\bibinfo{pages}{4074--4080}.
	\bibitem[{Novotny and Van~Hulst(2011)}]{novotny2011antennas}
	\bibinfo{author}{Novotny\xfnm[ L.]}, \bibinfo{author}{Van~Hulst\xfnm[ N.]}.
	\newblock \bibinfo{title}{Antennas for light}.
	\newblock \bibinfo{journal}{Nat Photonics}
	\bibinfo{year}{2011};\bibinfo{volume}{5}(\bibinfo{number}{2}):\bibinfo{pages}{83}.
	\bibitem[{Yang et~al.(2014)Yang, Kravchenko, Briggs and
		Valentine}]{yang2014all}
	\bibinfo{author}{Yang\xfnm[ Y.]}, \bibinfo{author}{Kravchenko\xfnm[ I.I.]},
	\bibinfo{author}{Briggs\xfnm[ D.P.]}, \bibinfo{author}{Valentine\xfnm[ J.]}.
	\newblock \bibinfo{title}{All-dielectric metasurface analogue of
		electromagnetically induced transparency}.
	\newblock \bibinfo{journal}{Nat Commun}
	\bibinfo{year}{2014};\bibinfo{volume}{5}:\bibinfo{pages}{5753}.
	\bibitem[{Zhang et~al.(2014{\natexlab{a}})Zhang, Liu, Zhu, Yuan and
		Qin}]{zhang2014strong}
	\bibinfo{author}{Zhang\xfnm[ J.]}, \bibinfo{author}{Liu\xfnm[ W.]},
	\bibinfo{author}{Zhu\xfnm[ Z.]}, \bibinfo{author}{Yuan\xfnm[ X.]},
	\bibinfo{author}{Qin\xfnm[ S.]}.
	\newblock \bibinfo{title}{Strong field enhancement and light-matter
		interactions with all-dielectric metamaterials based on split bar
		resonators}.
	\newblock \bibinfo{journal}{Opt Express}
	\bibinfo{year}{2014}{\natexlab{a}};\bibinfo{volume}{22}(\bibinfo{number}{25}):\bibinfo{pages}{30889--30898}.
	\bibitem[{Sun et~al.(2017)Sun, Yuan, Zhang, Zhang and Zhu}]{sun2017q}
	\bibinfo{author}{Sun\xfnm[ G.]}, \bibinfo{author}{Yuan\xfnm[ L.]},
	\bibinfo{author}{Zhang\xfnm[ Y.]}, \bibinfo{author}{Zhang\xfnm[ X.]},
	\bibinfo{author}{Zhu\xfnm[ Y.]}.
	\newblock \bibinfo{title}{Q-factor enhancement of fano resonance in
		all-dielectric metasurfaces by modulating meta-atom interactions}.
	\newblock \bibinfo{journal}{Sci Rep}
	\bibinfo{year}{2017};\bibinfo{volume}{7}(\bibinfo{number}{1}):\bibinfo{pages}{8128}.
	\bibitem[{Ding et~al.(2018)Ding, Yang, Deshpande and
		Bozhevolnyi}]{ding2018review}
	\bibinfo{author}{Ding\xfnm[ F.]}, \bibinfo{author}{Yang\xfnm[ Y.]},
	\bibinfo{author}{Deshpande\xfnm[ R.A.]}, \bibinfo{author}{Bozhevolnyi\xfnm[
		S.I.]}.
	\newblock \bibinfo{title}{A review of gap-surface plasmon metasurfaces:
		fundamentals and applications}.
	\newblock \bibinfo{journal}{Nanophotonics}
	\bibinfo{year}{2018};\bibinfo{volume}{7}(\bibinfo{number}{6}):\bibinfo{pages}{1129--1156}.
	\bibitem[{Zheludev and Kivshar(2012)}]{zheludev2012metamaterials}
	\bibinfo{author}{Zheludev\xfnm[ N.I.]}, \bibinfo{author}{Kivshar\xfnm[ Y.S.]}.
	\newblock \bibinfo{title}{From metamaterials to metadevices}.
	\newblock \bibinfo{journal}{Nat Mater}
	\bibinfo{year}{2012};\bibinfo{volume}{11}(\bibinfo{number}{11}):\bibinfo{pages}{917}.
	\bibitem[{Zhu et~al.(2011)Zhu, Liu, Zhang, Tsai, Bourouina, Teng
		et~al.}]{zhu2011switchable}
	\bibinfo{author}{Zhu\xfnm[ W.M.]}, \bibinfo{author}{Liu\xfnm[ A.Q.]},
	\bibinfo{author}{Zhang\xfnm[ X.M.]}, \bibinfo{author}{Tsai\xfnm[ D.P.]},
	\bibinfo{author}{Bourouina\xfnm[ T.]}, \bibinfo{author}{Teng\xfnm[ J.H.]},
	et~al.
	\newblock \bibinfo{title}{Switchable magnetic metamaterials using
		micromachining processes}.
	\newblock \bibinfo{journal}{Adv Mater}
	\bibinfo{year}{2011};\bibinfo{volume}{23}(\bibinfo{number}{15}):\bibinfo{pages}{1792--1796}.
	\bibitem[{Ou et~al.(2011)Ou, Plum, Jiang and Zheludev}]{ou2011reconfigurable}
	\bibinfo{author}{Ou\xfnm[ J.Y.]}, \bibinfo{author}{Plum\xfnm[ E.]},
	\bibinfo{author}{Jiang\xfnm[ L.]}, \bibinfo{author}{Zheludev\xfnm[ N.I.]}.
	\newblock \bibinfo{title}{Reconfigurable photonic metamaterials}.
	\newblock \bibinfo{journal}{Nano Lett}
	\bibinfo{year}{2011};\bibinfo{volume}{11}(\bibinfo{number}{5}):\bibinfo{pages}{2142--2144}.
	\bibitem[{Gu et~al.(2012)Gu, Singh, Liu, Zhang, Ma, Zhang
		et~al.}]{gu2012active}
	\bibinfo{author}{Gu\xfnm[ J.]}, \bibinfo{author}{Singh\xfnm[ R.]},
	\bibinfo{author}{Liu\xfnm[ X.]}, \bibinfo{author}{Zhang\xfnm[ X.]},
	\bibinfo{author}{Ma\xfnm[ Y.]}, \bibinfo{author}{Zhang\xfnm[ S.]}, et~al.
	\newblock \bibinfo{title}{Active control of electromagnetically induced
		transparency analogue in terahertz metamaterials}.
	\newblock \bibinfo{journal}{Nat Commun}
	\bibinfo{year}{2012};\bibinfo{volume}{3}:\bibinfo{pages}{1151}.
	\bibitem[{Gholipour et~al.(2013)Gholipour, Zhang, MacDonald, Hewak and
		Zheludev}]{gholipour2013all}
	\bibinfo{author}{Gholipour\xfnm[ B.]}, \bibinfo{author}{Zhang\xfnm[ J.]},
	\bibinfo{author}{MacDonald\xfnm[ K.F.]}, \bibinfo{author}{Hewak\xfnm[ D.W.]},
	\bibinfo{author}{Zheludev\xfnm[ N.I.]}.
	\newblock \bibinfo{title}{An all-optical, non-volatile, bidirectional,
		phase-change meta-switch}.
	\newblock \bibinfo{journal}{Adv Mater}
	\bibinfo{year}{2013};\bibinfo{volume}{25}(\bibinfo{number}{22}):\bibinfo{pages}{3050--3054}.
	\bibitem[{Xu et~al.(2016)Xu, Su, Ouyang, Xu, Cao, Zhang
		et~al.}]{xu2016frequency}
	\bibinfo{author}{Xu\xfnm[ Q.]}, \bibinfo{author}{Su\xfnm[ X.]},
	\bibinfo{author}{Ouyang\xfnm[ C.]}, \bibinfo{author}{Xu\xfnm[ N.]},
	\bibinfo{author}{Cao\xfnm[ W.]}, \bibinfo{author}{Zhang\xfnm[ Y.]}, et~al.
	\newblock \bibinfo{title}{Frequency-agile electromagnetically induced
		transparency analogue in terahertz metamaterials}.
	\newblock \bibinfo{journal}{Opt Lett}
	\bibinfo{year}{2016};\bibinfo{volume}{41}(\bibinfo{number}{19}):\bibinfo{pages}{4562--4565}.
	\bibitem[{Fan et~al.(2017)Fan, Qiao, Zhang, Fu, Dong, Kong
		et~al.}]{fan2017electromagnetic}
	\bibinfo{author}{Fan\xfnm[ Y.]}, \bibinfo{author}{Qiao\xfnm[ T.]},
	\bibinfo{author}{Zhang\xfnm[ F.]}, \bibinfo{author}{Fu\xfnm[ Q.]},
	\bibinfo{author}{Dong\xfnm[ J.]}, \bibinfo{author}{Kong\xfnm[ B.]}, et~al.
	\newblock \bibinfo{title}{An electromagnetic modulator based on electrically
		controllable metamaterial analogue to electromagnetically induced
		transparency}.
	\newblock \bibinfo{journal}{Sci Rep}
	\bibinfo{year}{2017};\bibinfo{volume}{7}:\bibinfo{pages}{40441}.
	\bibitem[{Zhang et~al.(2013{\natexlab{c}})Zhang, MacDonald and
		Zheludev}]{zhang2013nonlinear}
	\bibinfo{author}{Zhang\xfnm[ J.]}, \bibinfo{author}{MacDonald\xfnm[ K.F.]},
	\bibinfo{author}{Zheludev\xfnm[ N.I.]}.
	\newblock \bibinfo{title}{Nonlinear dielectric optomechanical metamaterials}.
	\newblock \bibinfo{journal}{Light Sci Appl}
	\bibinfo{year}{2013}{\natexlab{c}};\bibinfo{volume}{2}(\bibinfo{number}{8}):\bibinfo{pages}{e96}.
	\bibitem[{Zhang et~al.(2014{\natexlab{b}})Zhang, Chen, Wang, Zhao, He and
		Chen}]{zhang2014thermally}
	\bibinfo{author}{Zhang\xfnm[ F.]}, \bibinfo{author}{Chen\xfnm[ L.]},
	\bibinfo{author}{Wang\xfnm[ Y.]}, \bibinfo{author}{Zhao\xfnm[ Q.]},
	\bibinfo{author}{He\xfnm[ X.]}, \bibinfo{author}{Chen\xfnm[ K.]}.
	\newblock \bibinfo{title}{Thermally tunable electric mie resonance of
		dielectric cut-wire type metamaterial}.
	\newblock \bibinfo{journal}{Opt Express}
	\bibinfo{year}{2014}{\natexlab{b}};\bibinfo{volume}{22}(\bibinfo{number}{21}):\bibinfo{pages}{24908--24913}.
	\bibitem[{Iyer et~al.(2015)Iyer, Butakov and Schuller}]{iyer2015reconfigurable}
	\bibinfo{author}{Iyer\xfnm[ P.P.]}, \bibinfo{author}{Butakov\xfnm[ N.A.]},
	\bibinfo{author}{Schuller\xfnm[ J.A.]}.
	\newblock \bibinfo{title}{Reconfigurable semiconductor phased-array
		metasurfaces}.
	\newblock \bibinfo{journal}{ACS Photonics}
	\bibinfo{year}{2015};\bibinfo{volume}{2}(\bibinfo{number}{8}):\bibinfo{pages}{1077--1084}.
	\bibitem[{Rahmani et~al.(2017)Rahmani, Xu, Miroshnichenko, Komar,
		Camacho-Morales, Chen et~al.}]{rahmani2017reversible}
	\bibinfo{author}{Rahmani\xfnm[ M.]}, \bibinfo{author}{Xu\xfnm[ L.]},
	\bibinfo{author}{Miroshnichenko\xfnm[ A.E.]}, \bibinfo{author}{Komar\xfnm[
		A.]}, \bibinfo{author}{Camacho-Morales\xfnm[ R.]},
	\bibinfo{author}{Chen\xfnm[ H.]}, et~al.
	\newblock \bibinfo{title}{Reversible thermal tuning of all-dielectric
		metasurfaces}.
	\newblock \bibinfo{journal}{Adv Funct Mater}
	\bibinfo{year}{2017};\bibinfo{volume}{27}(\bibinfo{number}{31}):\bibinfo{pages}{1700580}.
	\bibitem[{Forouzmand et~al.(2018)Forouzmand, Salary, Inampudi and
		Mosallaei}]{forouzmand2018tunable}
	\bibinfo{author}{Forouzmand\xfnm[ A.]}, \bibinfo{author}{Salary\xfnm[ M.M.]},
	\bibinfo{author}{Inampudi\xfnm[ S.]}, \bibinfo{author}{Mosallaei\xfnm[ H.]}.
	\newblock \bibinfo{title}{A tunable multigate indium-tin-oxide-assisted
		all-dielectric metasurface}.
	\newblock \bibinfo{journal}{Adv Opt Mater}
	\bibinfo{year}{2018};\bibinfo{volume}{6}(\bibinfo{number}{7}):\bibinfo{pages}{1701275}.
	\bibitem[{Ju et~al.(2011)Ju, Geng, Horng, Girit, Martin, Hao
		et~al.}]{ju2011graphene}
	\bibinfo{author}{Ju\xfnm[ L.]}, \bibinfo{author}{Geng\xfnm[ B.]},
	\bibinfo{author}{Horng\xfnm[ J.]}, \bibinfo{author}{Girit\xfnm[ C.]},
	\bibinfo{author}{Martin\xfnm[ M.]}, \bibinfo{author}{Hao\xfnm[ Z.]}, et~al.
	\newblock \bibinfo{title}{Graphene plasmonics for tunable terahertz
		metamaterials}.
	\newblock \bibinfo{journal}{Nat Nanotechnol}
	\bibinfo{year}{2011};\bibinfo{volume}{6}(\bibinfo{number}{10}):\bibinfo{pages}{630}.
	\bibitem[{He(2015)}]{he2015tunable}
	\bibinfo{author}{He\xfnm[ X.]}.
	\newblock \bibinfo{title}{Tunable terahertz graphene metamaterials}.
	\newblock \bibinfo{journal}{Carbon}
	\bibinfo{year}{2015};\bibinfo{volume}{82}:\bibinfo{pages}{229--237}.
	\bibitem[{Li et~al.(2015)Li, Tian, Zhang, Xu, Singh, Gu et~al.}]{li2015dual}
	\bibinfo{author}{Li\xfnm[ Q.]}, \bibinfo{author}{Tian\xfnm[ Z.]},
	\bibinfo{author}{Zhang\xfnm[ X.]}, \bibinfo{author}{Xu\xfnm[ N.]},
	\bibinfo{author}{Singh\xfnm[ R.]}, \bibinfo{author}{Gu\xfnm[ J.]}, et~al.
	\newblock \bibinfo{title}{Dual control of active graphene--silicon hybrid
		metamaterial devices}.
	\newblock \bibinfo{journal}{Carbon}
	\bibinfo{year}{2015};\bibinfo{volume}{90}:\bibinfo{pages}{146--153}.
	\bibitem[{Xia et~al.(2018)Xia, Zhai, Wang and Wen}]{xia2018plasmonically}
	\bibinfo{author}{Xia\xfnm[ S.X.]}, \bibinfo{author}{Zhai\xfnm[ X.]},
	\bibinfo{author}{Wang\xfnm[ L.L.]}, \bibinfo{author}{Wen\xfnm[ S.C.]}.
	\newblock \bibinfo{title}{Plasmonically induced transparency in double-layered
		graphene nanoribbons}.
	\newblock \bibinfo{journal}{Photonics Res}
	\bibinfo{year}{2018};\bibinfo{volume}{6}(\bibinfo{number}{7}):\bibinfo{pages}{692--702}.
	\bibitem[{Farmani et~al.(2018)Farmani, Mir and Sharifpour}]{farmani2018broadly}
	\bibinfo{author}{Farmani\xfnm[ A.]}, \bibinfo{author}{Mir\xfnm[ A.]},
	\bibinfo{author}{Sharifpour\xfnm[ Z.]}.
	\newblock \bibinfo{title}{Broadly tunable and bidirectional terahertz graphene
		plasmonic switch based on enhanced goos-h{\"a}nchen effect}.
	\newblock \bibinfo{journal}{Appl Surf Sci}
	\bibinfo{year}{2018};\bibinfo{volume}{453}:\bibinfo{pages}{358--364}.
	\bibitem[{Li et~al.(2019{\natexlab{a}})Li, Ji, Ren, Hu, Qin and
		Wang}]{li2019investigation}
	\bibinfo{author}{Li\xfnm[ H.]}, \bibinfo{author}{Ji\xfnm[ C.]},
	\bibinfo{author}{Ren\xfnm[ Y.]}, \bibinfo{author}{Hu\xfnm[ J.]},
	\bibinfo{author}{Qin\xfnm[ M.]}, \bibinfo{author}{Wang\xfnm[ L.]}.
	\newblock \bibinfo{title}{Investigation of multiband plasmonic metamaterial
		perfect absorbers based on graphene ribbons by the phase-coupled method}.
	\newblock \bibinfo{journal}{Carbon}
	\bibinfo{year}{2019}{\natexlab{a}};\bibinfo{volume}{141}:\bibinfo{pages}{481--487}.
	\bibitem[{Yee et~al.(2011)Yee, Kim, Jung, Hong and Kong}]{yee2011ultrafast}
	\bibinfo{author}{Yee\xfnm[ K.J.]}, \bibinfo{author}{Kim\xfnm[ J.H.]},
	\bibinfo{author}{Jung\xfnm[ M.H.]}, \bibinfo{author}{Hong\xfnm[ B.H.]},
	\bibinfo{author}{Kong\xfnm[ K.J.]}.
	\newblock \bibinfo{title}{Ultrafast modulation of optical transitions in
		monolayer and multilayer graphene}.
	\newblock \bibinfo{journal}{Carbon}
	\bibinfo{year}{2011};\bibinfo{volume}{49}(\bibinfo{number}{14}):\bibinfo{pages}{4781--4785}.
	\bibitem[{Li et~al.(2014)Li, Chen, Meng, Fang, Xiao, Li
		et~al.}]{li2014ultrafast}
	\bibinfo{author}{Li\xfnm[ W.]}, \bibinfo{author}{Chen\xfnm[ B.]},
	\bibinfo{author}{Meng\xfnm[ C.]}, \bibinfo{author}{Fang\xfnm[ W.]},
	\bibinfo{author}{Xiao\xfnm[ Y.]}, \bibinfo{author}{Li\xfnm[ X.]}, et~al.
	\newblock \bibinfo{title}{Ultrafast all-optical graphene modulator}.
	\newblock \bibinfo{journal}{Nano Lett}
	\bibinfo{year}{2014};\bibinfo{volume}{14}(\bibinfo{number}{2}):\bibinfo{pages}{955--959}.
	\bibitem[{Li et~al.(2016)Li, Cong, Singh, Xu, Cao, Zhang
		et~al.}]{li2016monolayer}
	\bibinfo{author}{Li\xfnm[ Q.]}, \bibinfo{author}{Cong\xfnm[ L.]},
	\bibinfo{author}{Singh\xfnm[ R.]}, \bibinfo{author}{Xu\xfnm[ N.]},
	\bibinfo{author}{Cao\xfnm[ W.]}, \bibinfo{author}{Zhang\xfnm[ X.]}, et~al.
	\newblock \bibinfo{title}{Monolayer graphene sensing enabled by the strong
		fano-resonant metasurface}.
	\newblock \bibinfo{journal}{Nanoscale}
	\bibinfo{year}{2016};\bibinfo{volume}{8}(\bibinfo{number}{39}):\bibinfo{pages}{17278--17284}.
	\bibitem[{Xiao et~al.(2017)Xiao, Wang, Jiang, Yan, Cheng, Wang
		et~al.}]{xiao2017strong}
	\bibinfo{author}{Xiao\xfnm[ S.]}, \bibinfo{author}{Wang\xfnm[ T.]},
	\bibinfo{author}{Jiang\xfnm[ X.]}, \bibinfo{author}{Yan\xfnm[ X.]},
	\bibinfo{author}{Cheng\xfnm[ L.]}, \bibinfo{author}{Wang\xfnm[ B.]}, et~al.
	\newblock \bibinfo{title}{Strong interaction between graphene layer and fano
		resonance in terahertz metamaterials}.
	\newblock \bibinfo{journal}{J Phys D: Appl Phys}
	\bibinfo{year}{2017};\bibinfo{volume}{50}(\bibinfo{number}{19}):\bibinfo{pages}{195101}.
	\bibitem[{Chen and Fan(2017)}]{chen2017study}
	\bibinfo{author}{Chen\xfnm[ X.]}, \bibinfo{author}{Fan\xfnm[ W.]}.
	\newblock \bibinfo{title}{Study of the interaction between graphene and planar
		terahertz metamaterial with toroidal dipolar resonance}.
	\newblock \bibinfo{journal}{Opt Lett}
	\bibinfo{year}{2017};\bibinfo{volume}{42}(\bibinfo{number}{10}):\bibinfo{pages}{2034--2037}.
	\bibitem[{Xiao et~al.(2018)Xiao, Wang, Liu, Yan, Li and Xu}]{xiao2018active}
	\bibinfo{author}{Xiao\xfnm[ S.]}, \bibinfo{author}{Wang\xfnm[ T.]},
	\bibinfo{author}{Liu\xfnm[ T.]}, \bibinfo{author}{Yan\xfnm[ X.]},
	\bibinfo{author}{Li\xfnm[ Z.]}, \bibinfo{author}{Xu\xfnm[ C.]}.
	\newblock \bibinfo{title}{Active modulation of electromagnetically induced
		transparency analogue in terahertz hybrid metal-graphene metamaterials}.
	\newblock \bibinfo{journal}{Carbon}
	\bibinfo{year}{2018};\bibinfo{volume}{126}:\bibinfo{pages}{271--278}.
	\bibitem[{Kim et~al.(2018)Kim, Kim, Zhao, Oh, Ha, Chung
		et~al.}]{kim2018electrically}
	\bibinfo{author}{Kim\xfnm[ T.T.]}, \bibinfo{author}{Kim\xfnm[ H.D.]},
	\bibinfo{author}{Zhao\xfnm[ R.]}, \bibinfo{author}{Oh\xfnm[ S.S.]},
	\bibinfo{author}{Ha\xfnm[ T.]}, \bibinfo{author}{Chung\xfnm[ D.S.]}, et~al.
	\newblock \bibinfo{title}{Electrically tunable slow light using graphene
		metamaterials}.
	\newblock \bibinfo{journal}{ACS Photonics}
	\bibinfo{year}{2018};\bibinfo{volume}{5}(\bibinfo{number}{5}):\bibinfo{pages}{1800--1807}.
	\bibitem[{Li et~al.(2019{\natexlab{b}})Li, Nugraha, Su, Chen, Yang, Unferdorben
		et~al.}]{li2019terahertz}
	\bibinfo{author}{Li\xfnm[ S.]}, \bibinfo{author}{Nugraha\xfnm[ P.S.]},
	\bibinfo{author}{Su\xfnm[ X.]}, \bibinfo{author}{Chen\xfnm[ X.]},
	\bibinfo{author}{Yang\xfnm[ Q.]}, \bibinfo{author}{Unferdorben\xfnm[ M.]},
	et~al.
	\newblock \bibinfo{title}{Terahertz electric field modulated mode coupling in
		graphene-metal hybrid metamaterials}.
	\newblock \bibinfo{journal}{Opt Express}
	\bibinfo{year}{2019}{\natexlab{b}};\bibinfo{volume}{27}(\bibinfo{number}{3}):\bibinfo{pages}{2317--2326}.
	\bibitem[{Argyropoulos(2015)}]{argyropoulos2015enhanced}
	\bibinfo{author}{Argyropoulos\xfnm[ C.]}.
	\newblock \bibinfo{title}{Enhanced transmission modulation based on dielectric
		metasurfaces loaded with graphene}.
	\newblock \bibinfo{journal}{Opt Express}
	\bibinfo{year}{2015};\bibinfo{volume}{23}(\bibinfo{number}{18}):\bibinfo{pages}{23787--23797}.
	\bibitem[{Chen et~al.(2017)Chen, Argyropoulos, Farhat and
		Gomez-Diaz}]{chen2017flatland}
	\bibinfo{author}{Chen\xfnm[ P.Y.]}, \bibinfo{author}{Argyropoulos\xfnm[ C.]},
	\bibinfo{author}{Farhat\xfnm[ M.]}, \bibinfo{author}{Gomez-Diaz\xfnm[ J.S.]}.
	\newblock \bibinfo{title}{Flatland plasmonics and nanophotonics based on
		graphene and beyond}.
	\newblock \bibinfo{journal}{Nanophotonics}
	\bibinfo{year}{2017};\bibinfo{volume}{6}(\bibinfo{number}{6}):\bibinfo{pages}{1239--1262}.
	\bibitem[{Suk et~al.(2011)Suk, Kitt, Magnuson, Hao, Ahmed, An
		et~al.}]{suk2011transfer}
	\bibinfo{author}{Suk\xfnm[ J.W.]}, \bibinfo{author}{Kitt\xfnm[ A.]},
	\bibinfo{author}{Magnuson\xfnm[ C.W.]}, \bibinfo{author}{Hao\xfnm[ Y.]},
	\bibinfo{author}{Ahmed\xfnm[ S.]}, \bibinfo{author}{An\xfnm[ J.]}, et~al.
	\newblock \bibinfo{title}{Transfer of cvd-grown monolayer graphene onto
		arbitrary substrates}.
	\newblock \bibinfo{journal}{ACS Nano}
	\bibinfo{year}{2011};\bibinfo{volume}{5}(\bibinfo{number}{9}):\bibinfo{pages}{6916--6924}.
	\bibitem[{Palik and Ghosh(1985)}]{palik1985handbook}
	\bibinfo{author}{Palik\xfnm[ E.]}, \bibinfo{author}{Ghosh\xfnm[ G.]}.
	\newblock \bibinfo{title}{Handbook of Optical Constants of Solids (Orlando, FL:
		Academic)}.
	\newblock \bibinfo{year}{1985}.
	\bibitem[{Zhang et~al.(2015)Zhang, Zhu, Liu, Yuan and Qin}]{zhang2015towards}
	\bibinfo{author}{Zhang\xfnm[ J.]}, \bibinfo{author}{Zhu\xfnm[ Z.]},
	\bibinfo{author}{Liu\xfnm[ W.]}, \bibinfo{author}{Yuan\xfnm[ X.]},
	\bibinfo{author}{Qin\xfnm[ S.]}.
	\newblock \bibinfo{title}{Towards photodetection with high efficiency and
		tunable spectral selectivity: graphene plasmonics for light trapping and
		absorption engineering}.
	\newblock \bibinfo{journal}{Nanoscale}
	\bibinfo{year}{2015};\bibinfo{volume}{7}(\bibinfo{number}{32}):\bibinfo{pages}{13530--13536}.
	\bibitem[{Xiao et~al.(2016)Xiao, Wang, Liu, Xu, Han and Yan}]{xiao2016tunable}
	\bibinfo{author}{Xiao\xfnm[ S.]}, \bibinfo{author}{Wang\xfnm[ T.]},
	\bibinfo{author}{Liu\xfnm[ Y.]}, \bibinfo{author}{Xu\xfnm[ C.]},
	\bibinfo{author}{Han\xfnm[ X.]}, \bibinfo{author}{Yan\xfnm[ X.]}.
	\newblock \bibinfo{title}{Tunable light trapping and absorption enhancement
		with graphene ring arrays}.
	\newblock \bibinfo{journal}{Phys Chem Chem Phys}
	\bibinfo{year}{2016};\bibinfo{volume}{18}(\bibinfo{number}{38}):\bibinfo{pages}{26661--26669}.
	\bibitem[{Kim et~al.(2012)Kim, Son, Cho, Geng, Regan, Shi
		et~al.}]{kim2012electrical}
	\bibinfo{author}{Kim\xfnm[ J.]}, \bibinfo{author}{Son\xfnm[ H.]},
	\bibinfo{author}{Cho\xfnm[ D.J.]}, \bibinfo{author}{Geng\xfnm[ B.]},
	\bibinfo{author}{Regan\xfnm[ W.]}, \bibinfo{author}{Shi\xfnm[ S.]}, et~al.
	\newblock \bibinfo{title}{Electrical control of optical plasmon resonance with
		graphene}.
	\newblock \bibinfo{journal}{Nano Lett}
	\bibinfo{year}{2012};\bibinfo{volume}{12}(\bibinfo{number}{11}):\bibinfo{pages}{5598--5602}.
	\bibitem[{Wu et~al.(2014)Wu, Arju, Kelp, Fan, Dominguez, Gonzales
		et~al.}]{wu2014spectrally}
	\bibinfo{author}{Wu\xfnm[ C.]}, \bibinfo{author}{Arju\xfnm[ N.]},
	\bibinfo{author}{Kelp\xfnm[ G.]}, \bibinfo{author}{Fan\xfnm[ J.A.]},
	\bibinfo{author}{Dominguez\xfnm[ J.]}, \bibinfo{author}{Gonzales\xfnm[ E.]},
	et~al.
	\newblock \bibinfo{title}{Spectrally selective chiral silicon metasurfaces
		based on infrared fano resonances}.
	\newblock \bibinfo{journal}{Nat Commun}
	\bibinfo{year}{2014};\bibinfo{volume}{5}:\bibinfo{pages}{3892}.
	\bibitem[{Hass et~al.(2008)Hass, Varchon, Millan-Otoya, Sprinkle, Sharma,
		de~Heer et~al.}]{hass2008multilayer}
	\bibinfo{author}{Hass\xfnm[ J.]}, \bibinfo{author}{Varchon\xfnm[ F.]},
	\bibinfo{author}{Millan-Otoya\xfnm[ J.E.]}, \bibinfo{author}{Sprinkle\xfnm[
		M.]}, \bibinfo{author}{Sharma\xfnm[ N.]}, \bibinfo{author}{de~Heer\xfnm[
		W.A.]}, et~al.
	\newblock \bibinfo{title}{Why multilayer graphene on 4 h- sic (000 1) behaves
		like a single sheet of graphene}.
	\newblock \bibinfo{journal}{Phys Rev Lett}
	\bibinfo{year}{2008};\bibinfo{volume}{100}(\bibinfo{number}{12}):\bibinfo{pages}{125504}.
	
\end{thebibliography}
\end{document}